\documentclass[twocolumn]{aastex62}
\usepackage{amsmath,amstext}
\usepackage{apjfonts} 
\usepackage{chngcntr}

\newcommand{\be}{\begin{eqnarray}}
\newcommand{\ee}{\end{eqnarray}}
\newcommand{\lp}{\left(}
\newcommand{\rp}{\right)}
\newcommand{\lb}{\left[}
\newcommand{\rb}{\right]}

\shorttitle{Wind-reprocessed Transients}
\shortauthors{Piro \& Lu}

\begin{document}

\title{\Large \textbf{Wind-reprocessed Transients}}

\author{Anthony L. Piro}
\affiliation{The Observatories of the Carnegie Institution for Science, 813 Santa Barbara St., Pasadena, CA 91101, USA; piro@carnegiescience.edu}
\author{Wenbin Lu}
\affiliation{Theoretical Astrophysics, and Walter Burke Institute for Theoretical Physics, Mail Code
350-17, Caltech, Pasadena, CA 91125, USA; wenbinlu@caltech.edu}

\begin{abstract}
We consider the situation where the luminosity from a transient event is reprocessed by an optically thick wind. Potential applications are the tidal disruption of stars by black holes, engine-powered supernovae, and unique fast transients found by current and future wide-field surveys. We derive relations between the injected and observed luminosity for steady and time dependent winds, and discuss how the temperature is set for scattering-dominated radiative transport. We apply this framework to specific examples of tidal disruption events and the formation of a black hole by a massive star, as well as discuss  other applications such as deriving observables from detailed hydrodynamic simulations.
We conclude by exploring what is inferred about the mass loss rate and underlying engine powering AT2018cow if it is explained as a wind-reprocessed transient, demonstrating that its optical emission is consistent with reprocessing of the observed soft X-rays.
\end{abstract}


\section{Introduction}

With the growth of wide-field and high-cadence surveys in recent years \citep[e.g.,][]{Brown13,Shappee14,Chambers16,Tartaglia18,Tonry18,Graham19}, the study of astrophysical transients has literally exploded. This has led to increasingly detailed studies of well-known transients (e.g., thermonuclear and core-collapse supernovae, classical novae, and gamma-ray bursts) as well as the almost regular discovery and study of a vast range of new transients, including tidal disruption events \citep[TDEs;][]{2012Natur.485..217G,Holoien14}, kilonovae \citep{Coulter17}, fast blue transients \citep[FBOTs;][]{Drout14}, calcium-rich transients \citep{Kasliwal12}, fast radio bursts \citep[FRBs;][]{Lorimer07,Thornton13}, and luminous red novae \citep{Rau07,Kasliwal11,Williams15}, just to name a few.

These new events have in turn inspired astrophysicists to consider novel methods to power them, such as shock interaction \citep{Balberg11,Chevalier11}, fallback accretion \citep{Dexter13}, radioactive heating from sources other than $^{56}$Ni \citep{Metzger10}, and spin down of highly-magnetized neutron stars \citep{Kasen10}. Besides the underlying energy source, a critical aspect for determining the observed properties of transients is the direct local environment around them. Perhaps no where is this better exemplified than with studies of interacting supernovae (e.g., Type IIn), where the properties can vary dramatically depending on the surrounding material \citep[e.g.,][]{Smith07,Ofek13}, or TDEs, where the optical emission is likely the result of an underlying powering source being reprocessed \citep[e.g.,][]{Strubbe09,Miller15,Roth16,Metzger16, Dai18, 2020MNRAS.492..686L}.

Motivated by these issues, we present a theoretical study of how the observed properties of a transient are altered when reprocessed by an outflow (or wind). Some of the basic framework for such a model was initially presented in the work of \citet{Strubbe09} in the specific context of TDEs. Here we take a  more general point of view so that in the future a broad range of powering sources and mass loss rates can be considered depending on the specific system of interest. Thus as new transients are discovered, the models here can help investigate whether a wind-reprocessed transient is a possible explanation, and if so, what it implies about the systems. Alternatively, this framework can be applied to specific theoretical models to make observational predictions. This could be especially useful as a way to post-process detailed hydrodynamic simulations to predict observables that would be too expensive to calculate using full radiative transfer.

In Section \ref{sec:steady}, we begin by considering the case of a steady (i.e., time-independent mass loading factor) wind. This helps provide some of the basic physical intuition for more complicated cases presented later. In Section \ref{sec:evolving}, we consider how the situation is modified if the wind can now change with time. In Section \ref{sec:temperature}, we investigate how the temperature of the reprocessed emission is expected to evolve, highlighting the importance of scattering-dominated radiative transport. In Section \ref{sec:examples}, we consider toy models of TDEs and black hole (BH) formation in the context of our framework, and then in Section \ref{sec:18cow} we discuss what is implied about AT2018cow if it is explained as a wind-reprocessed transient. We conclude in Section \ref{sec:conclusions} with a summary of our results and a discussion of future work.

\section{Steady Wind}
\label{sec:steady}

\begin{figure}
\includegraphics[width=0.45\textwidth,trim=0.0cm 0.0cm 0.0cm 0.0cm]{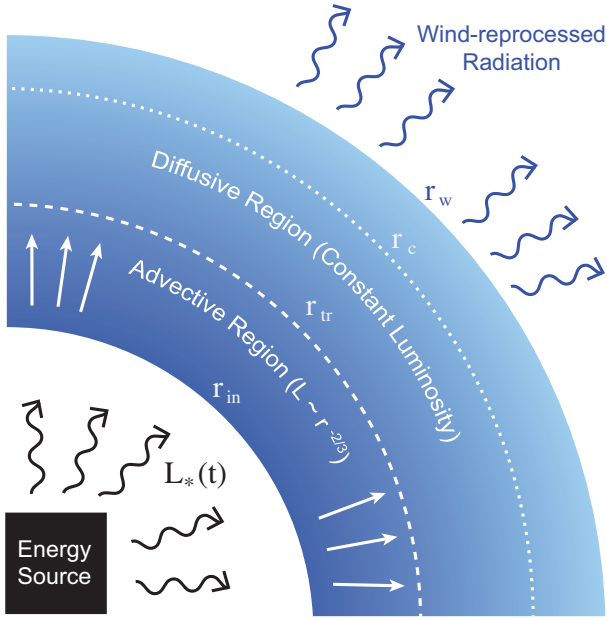}
\caption{Diagram highlighting the key regions for a wind-reprocessed transient. An energy source (denoted by a black box) injects a luminosity $L_*(t)$, which heats a wind (denoted by the blue region) at a radius $r_{\rm in}$. In the dense inner regions of the wind, this luminosity is advected along with the wind material out to the trapping radius $r_{\rm tr}$ (marked with a dashed line), above which the luminosity is roughly constant. Within this outer region, the outermost radius where thermalization can occur is at the color radius $r_c$ (marked with a dotted line). The radii $r_{\rm tr}$ and $r_c$ thus determine the luminosity and temperature, respectively, of the wind-reprocessed radiation seen by an observed.}
\label{fig:diagram}
\end{figure}

For the basic setup, as shown in Figure \ref{fig:diagram}, consider a luminosity $L_*$ that is reprocessed by a steady, optically-thick wind with velocity $v_w$ and mass-loss rate $\dot{M}$. The density profile of the wind is set by mass continuity to be
\be
    \rho(r) = \frac{\dot{M}}{4\pi r^2 v_w }= \frac{K}{r^2},
\ee
where $K$ is the mass-loading factor. The outer boundary of the wind evolves with time as
\be
    r_w = r_{\rm in} + v_w t,
\ee
where $r_{\rm in}$ is the wind's inner boundary.

Given the temperatures and densities present for the scenarios we will be considering, the opacity of the wind is generally dominated by electron scattering $\kappa_s=0.34\,\rm cm^2\,g^{-1}$ (for a solar-like composition). Note that the absorption opacity is still important for determining the observed color temperature of the transient, which we address in more detail in Section \ref{sec:temperature}. The scattering optical depth is given by
\be
    \tau(r) = \int_r^{r_w} \kappa_s \rho dr = \kappa_s K(r^{-1}-r_w^{-1}).
    \label{eq:tau1}
\ee
The photon diffusion time at a radius $r<r_w$ can be estimated as
\be
    t_{\rm dif} \approx \frac{\tau(r)}{c}\frac{(r_w-r)r}{r_w},
    \label{eq:tdif}
\ee
which matches the expected limits of $t_{\rm dif}\approx (r_w-r)\tau/c$ when $r\approx r_w$ and $t_{\rm dif}\approx r\tau/c$ when $r\ll r_w$. The dynamical time of a shell at radius $r$ is
\be
    t_{\rm dyn} \approx (r-r_{\rm in})/v_w.
    \label{eq:tdyn}
\ee
The photon trapping radius $r_{\rm tr}$ is defined as the depth where $t_{\rm dif}=t_{\rm dyn}$, which results in an algebraic expression
\be
    \frac{r_{\rm tr}}{r_{\rm in}}
        = 1 + \frac{\kappa_s K}{r_{\rm in}}\frac{v_w}{c}
            \frac{(r_w-r_{\rm tr})^2}{r_w^2}
        = 1 + A\frac{(r_w-r_{\rm tr})^2}{r_w^2}.
    \label{eq:rtr}
\ee
The dimensionless constant
\be
    A\equiv \kappa_s Kv_w/r_{\rm in}c,
\ee
reflects how strongly radiation is trapped at the inner radius $r_{\rm in}$. Thus, we require $A\gtrsim 1$, otherwise photons are never trapped. Equation~(\ref{eq:rtr}) is quadratic in $r_{\rm tr}$, and so it can be easily solved for given values of $r_{\rm in}$, $r_w$, and $A$.

Photons injected at the inner radius $r_{\rm in}$ with luminosity $L_*$ are advected along with the wind out to the trapping radius $r_{\rm tr}$. This causes them to be adiabatically degraded, so that their energy density scales $\mathcal{E}\propto \rho^{4/3} \propto r^{-8/3}$ \citep{Strubbe09}. Below the trapping depth the energy density is then
\be
    \mathcal{E}(r)
    = \frac{L_*}{4\pi r_{\rm in}^2 v_w}
        \lp \frac{r}{r_{\rm in}} \rp^{-8/3}.
\ee
Above the trapping depth, there is little adiabatic cooling and the luminosity is roughly constant. The observed luminosity is set by the flux of radiation across the trapping depth
\be
    L_{\rm obs} = 4\pi r_{\rm tr}^2\mathcal{E}(r_{\rm tr})
    \lp v_w - \frac{dr_{\rm tr}}{dt} \rp,
\ee
where note we have been careful to include the effect of the changing trapping depth (reflected in the $dr_{\rm tr}/dt$ term) that is often not included in other similar analytic treatments. Thus, the ratio between the observed and injected luminosities is
\be
    \frac{L_{\rm obs}}{L_*}
    = \lp \frac{r_{\rm tr}}{r_{\rm in}} \rp^{-2/3}
        \lp 1 - \frac{1}{v_w}\frac{dr_{\rm tr}}{dt} \rp,
    \label{eq:lobs}
\ee
Combining Equations~(\ref{eq:rtr}) and (\ref{eq:lobs}) provides the full time evolution of the observed luminosity. We next analytically estimate the evolution of the ratio $L_{\rm obs}/L_*$ at different characteristic times during the expansion of the wind.

\subsection{Early Times}

The ratio of the wind expansion to its initial radius gives a dimensionless measure of the time $v_wt/r_{\rm in}$. For $v_wt/r_{\rm in}\ll1$, the wind has not expanded appreciably and $r_{\rm tr}\approx r_w \sim r_{\rm in}$. Evaluating Equation~(\ref{eq:rtr}) in this limit,
\be
    (r_w-r_{\rm tr})^2
    = \frac{r_w^2}{A} \frac{r_{\rm tr}-r_{\rm in}}{r_{\rm in}}
    \approx \frac{r_w^2}{A}\frac{r_{\rm w}-r_{\rm in}}{r_{\rm in}}
    = \frac{r_w^2v_wt}{Ar_{\rm in}}.
\ee
Taking the square root of this expression and then the time derivative, we obtain
\be
    1-\frac{1}{v_w}\frac{dr_{\rm tr}}{dt}
    \approx \frac{1}{2}\lp \frac{r_w^2}{Ar_{\rm in} v_w t} \rp^{1/2}.
\ee
Substituting this into Equation (\ref{eq:lobs}),
\be
    \frac{L_{\rm obs}}{L_*}
    \approx  \frac{1}{2}\lp \frac{r_{\rm tr}}{r_{\rm in}} \rp^{-2/3}
        \lp \frac{r_w}{Ar_{\rm in}} \rp^{1/2} \lp \frac{r_w}{v_w t} \rp^{1/2}.
\ee
Thus as sufficiently early times $L_{\rm obs}/L_*\propto t^{-1/2}$.

\subsection{Middle Times}

Next, for $v_wt/r_{\rm in}\gg1$, then $r_{\rm tr}\approx r_w \approx v_w t \gg r_{\rm in}$, and from Equation~(\ref{eq:rtr}) we find,
\be
    (r_w-r_{\rm tr})^2
    = \frac{r_w^2}{A} \frac{r_{\rm tr}-r_{\rm in}}{r_{\rm in}}
    \approx \frac{r_w^3}{Ar_{\rm in}}
    \approx \frac{v_w^3t^3}{Ar_{\rm in}}.
\ee
Again taking the square root of this expression and then the time derivative,
\be
    1-\frac{1}{v_w}\frac{dr_{\rm tr}}{dt}
    \approx \frac{3}{2}\lp \frac{v_wt}{Ar_{\rm in}}\rp^{1/2}.
\ee
so that
\be
    \frac{L_{\rm obs}}{L_*}
        \approx 
        \frac{3}{2}
            \lp \frac{r_{\rm tr}}{r_{\rm in}} \rp^{-2/3}
            \lp \frac{v_w t}{Ar_{\rm in}} \rp^{1/2}.
\ee
Now that $r_{\rm rt}\approx v_wt$, then $L_{\rm obs}/L_*\propto t^{-2/3}t^{1/2}\propto t^{-1/6}$.

\subsection{Late Times}
\label{sec:late times}

Eventually, $v_w t/r_{\rm in}\gg A$, so that $r_w\gg r_{\rm tr}\gg r_{\rm in}$. In this case Equation (\ref{eq:rtr}) can be used to show $r_{\rm tr}\approx Ar_{\rm in}$, which is constant with time. Thus $1-v_w^{-1}dr_{\rm tr}/dt\approx 1$. This is equivalent to taking $\tau(r_{\rm tr})\approx c/v_w$, which is the classic condition typically used for the trapping radius \citep[e.g.,][]{Strubbe09}. Here we see that this is only valid when the wind has expanded sufficiently away from the trapping radius, and that the solution to Equation (\ref{eq:rtr}) should be used in general. In this case,
\be
    \frac{L_{\rm obs}}{L_*}
        \approx  A^{-2/3},
\ee
which is just a fixed ratio with time as long as $A$ is constant (we consider an evolving $A$ in Section \ref{sec:evolving}).

\subsection{Full Solutions for Steady Wind}

\begin{figure}
\includegraphics[width=0.48\textwidth,trim=0.0cm 0.0cm 0.0cm 0.5cm]{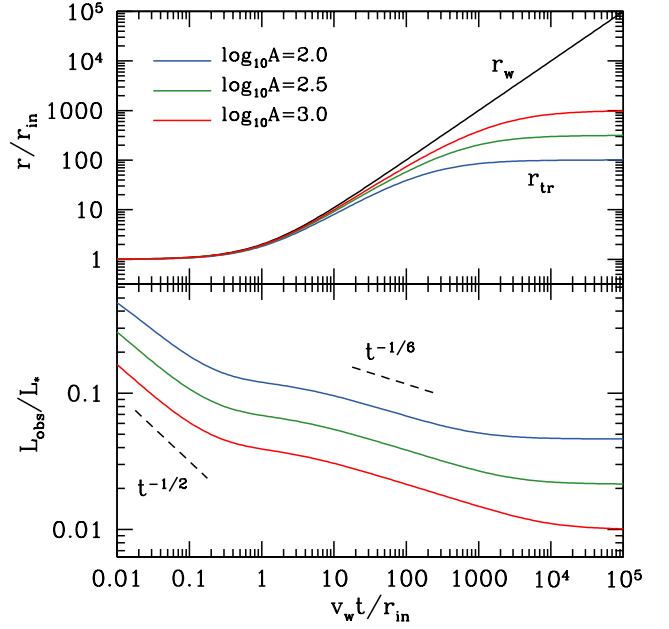}
\caption{Radius and luminosity evolution for a steady wind. The upper panel plots the wind radius $r_w$ (black line) and the trapping radius $r_{\rm tr}$, which is solved from Equation (\ref{eq:rtr}), for different values of $A$ as indicated. The trapping radius evolution shows distinct changes at $v_wt/r_{\rm in}\approx1$ and $v_wt/r_{\rm in}\approx A$, which correspond to breaks between the early, middle, and late stages. The bottom panel plots the observed luminosity ratio using Equation (\ref{eq:lobs}). This exhibits power-law evolution as derived in the text.}
\label{fig:steady}
\end{figure}

Full solutions for the evolution of a steady wind are plotted in Figure \ref{fig:steady} for different values of $A$. This shows the expected features estimated analytically in the previous sections. The trapping radius initially evolves along with $r_w$, but then asymptotes to $r_{\rm tr}/r_{\rm in}\approx A$ once $v_w t/r_{\rm in}\gg A$. The luminosity evolves from $L_{\rm obs}\propto t^{-1/2}$ to $L_{\rm obs}\propto t^{-1/6}$ before finally asymptoting to $L_{\rm obs}/L_*= A^{-2/3}$.

We caution though that some of the features of this evolution are more for academic interest. Early on the wind may require a timescale $\sim r_{\rm in}/v_w$ to develop. Furthermore, there may be a timescale associated with actually generating the illuminating luminosity. For these reasons, there will likely be a rise to peak that is not resolved in the treatment here.

\section{Evolving Wind}
\label{sec:evolving}

In general, one might expect the mass loss of the wind to evolve, so we next consider the more general case where the mass loading parameter $K$ is a function of time.

\subsection{Basic Framework}
\label{sec:basic}

We assume that the velocity of the wind keeps a characteristic constant value $v_w$. If a shell is launched into the wind at a radius $r_{\rm in}$ and at a time $t_0$, then it reaches a radius $r$ at a time
\be
    t = t_0 + (r-r_{\rm in})/v_w.
        \label{eq:launchingtime}
\ee
This means that if we want the density profile at any time $t$, then it is given by
\be
    \rho(r,t) = K[t_0(r,t)]/r^2,
\ee
and the optical depth at any time and radius is
\be
    \tau(r,t) = \int_r^{r_w} \kappa_s \rho(r,t) dr
    = \kappa_s  \int_r^{r_w} \frac{K[t_0(r,t)]}{r^2} dr.
        \label{eq:tau2}
\ee
The trapping radius is found by equating the diffusion and dynamical times
\be
    \frac{\tau(r_{\rm tr})}{c}\frac{(r_w-r_{\rm tr})r_{\rm tr}}{r_w} = t-t_0.
    \label{eq:rtr_condition}
\ee
The observed luminosity is
\be
    L_{\rm obs}(t)
    = L_*[t_0(r_{\rm tr},t)] \lb\frac{r_{\rm tr}(t)}{r_{\rm in}} \rb^{-2/3}
     \lp 1-\frac{1}{v_w}\frac{dr_{\rm tr}}{dt} \rp,
\ee
where we note one must be careful to evaluate the injected luminosity $L_*$ at the injection timescale $t_0$ for the current trapping radius.

\subsection{Power-Law Wind Evolution}
\label{sec:power law}

To provide more intuition about how the an evolving wind differs from a steady wind, it is helpful to consider some toy models. In the simplest physically-motivated cases, such as winds driven from a disk or fallback in a tidal disruption event, the wind-loading factor scales as a power law with time with the form
\be
    K(t) = K_{\rm max}(1+t/t')^{-\beta},
    \label{eq:power law k}
\ee
where $K_{\rm max}$ is the maximum wind-loading factor, $t'$ is the timescale for the wind to begin changing, and a typical value for the power law is $\beta\approx5/3$ (other values such as $\beta=4/3$ may be considered for a wind driven from a radiatively inefficient disk). Rewritten in dimensionless terms, the wind loading can be expressed as
\be
    A(t) = A_{\rm max}(1+t/t')^{-\beta}
\ee
where $A_{\rm max} \equiv \kappa_s K_{\rm max} v_w/r_{\rm in}c$. The introduction of the additional timescale $t'$, in comparison to the previously discussed timescale $r_{\rm in}/v_w$, results in a richer diversity of evolution for $L_{\rm obs}/L_*$.

We next solve the set of equations described above for power-law wind evolution. Here we just summarize how solutions for a time-dependent wind differ from a steady one. Readers interested in the specific techniques we employ to calculate these solutions should consult Appendix~\ref{sec:solving}.

\begin{figure}
\includegraphics[width=0.48\textwidth,trim=0.0cm 0.0cm 0.0cm 0.5cm]{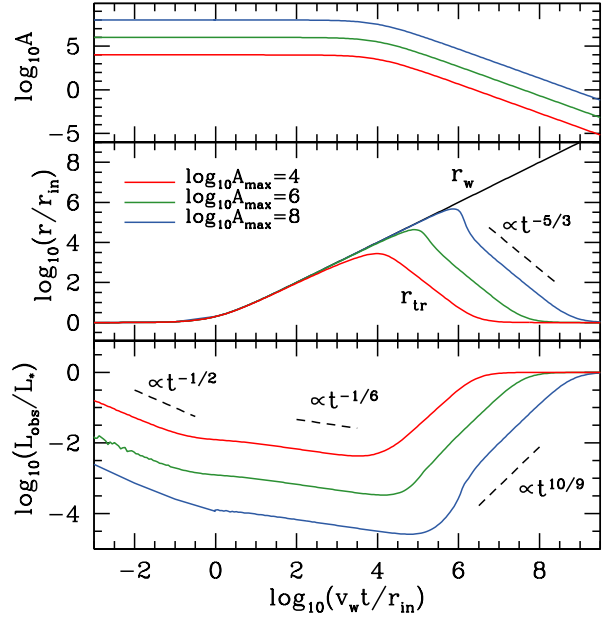}
\caption{The evolution for $v_wt'/r_{\rm in}=10^4$ and $\beta=5/3$ with different values of $A_{\rm max}$ as denoted. For $t<t'$, the wind is constant and mostly matches our steady wind solutions. For $t>t'$, the wind mass loss decreases and $r_{\rm tr}$ and $L_{\rm obs}$ follow the power law evolution given by Equations~(\ref{eq:rtr late}) and (\ref{eq:lobs late}), respectively.}
\label{fig:fallback_general}
\end{figure}

First, in Figure \ref{fig:fallback_general} we consider how the solutions evolve with different values of $A_{\rm max}$ and the dimensionless wind time is relatively large in comparison to unity with $v_wt'/r_{\rm in}=10^4$. For times $t<t'$, the wind is roughly constant and the evolution goes through two stages that roughly mimic what we found before for the steady wind case. Namely, initially the trapping radius is roughly constant with $L_{\rm obs}/L_*\propto t^{-1/2}$ and next $r_{\rm tr}\approx r_w\propto t$ with $L_{\rm obs}/L_*\propto t^{-1/6}$.

For times $t>t'$, the wind is changing with $A(t)\approx A_{\rm max}(t/t')^{-\beta}$. Using the late time solutions from Section~\ref{sec:late times}, the trapping radius then evolves as
\be
    r_{\rm tr} \approx Ar_{\rm in}
        \approx A_{\rm max}r_{\rm in}(t/t')^{-\beta},
    \label{eq:rtr late}
\ee
and the observed luminosity as
\be
    L_{\rm obs}/L_*\approx A^{-2/3}
    \approx A_{\rm max}^{-2/3}(t/t')^{2\beta/3}.
    \label{eq:lobs late}
\ee
Both scalings match what we find numerically for $\beta=5/3$. Note though that the changes in the evolution of $r_{\rm tr}$ and $L_{\rm obs}$ happen later than the change in $A$. This is because if the wind evolution changes at a time $t'$ then the trapping radius and luminosity only react at a later time of $\approx t'+(r_{\rm tr}-r_w)/v_w$. Eventually $r_{\rm tr}\approx r_{\rm in}$ and stops evolving with $L_{\rm obs}\approx L_*$. This happens when
\be
    t>t'A_{\rm max}^{1/\beta}.
\ee
This occurs later for larger $t'$ and $A_{\rm max}$ as also shown by the numerical solutions.

\begin{figure}
\includegraphics[width=0.48\textwidth,trim=0.0cm 0.0cm 0.0cm 0.5cm]{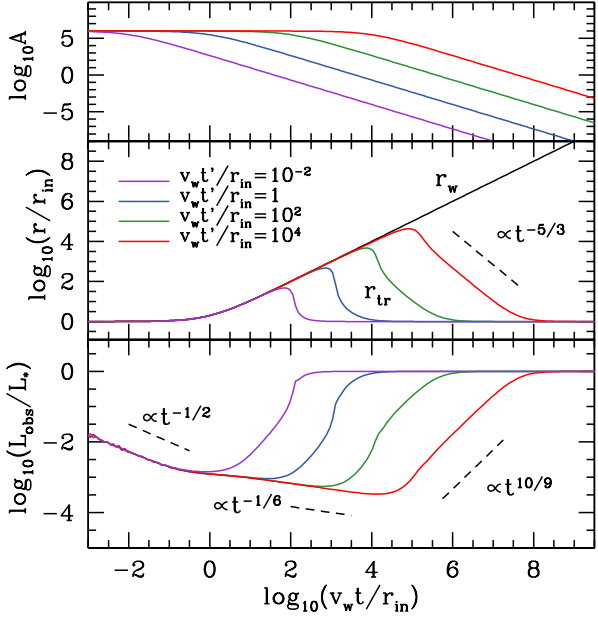}
\caption{The evolution when we fix $A_{\rm max}=10^6$ and $\beta=5/3$ and vary $t'$. For large values of $t'$, the evolution matches what was found in Figure \ref{fig:fallback_general}. As we decrease $t'$, the intermediate stage (where $r_{\rm tr}\approx r_w$ and $L_{\rm obs}/L_*\propto t^{-1/6}$) gets shorter and shorter. Finally, for \mbox{$v_wt'/r_{\rm in}\lesssim1$} (purple line), the luminosity transitions directly between early and late phases.}
\label{fig:fallback_general2}
\end{figure}

In Figure \ref{fig:fallback_general2}, we now fix $A_{\rm max}=10^6$ and instead vary the value of $t'$. When \mbox{$v_wt'/r_{\rm in}\gg1$,} then the evolution matches what was found above. For smaller values of $t'$, the solutions transition sooner to the phase where the trapping radius is moving back into the wind. Note though that because the solutions also transition sooner to $r_{\rm tr}\approx r_{\rm in}$, the power laws of $r_{\rm tr}\propto t^{-\beta}$ and $L_{\rm obs}/L_*\propto t^{2\beta/3}$ are not obeyed as closely. Thus we expect in practice that when $t'$ is small it will be more difficult to infer exactly what $\beta$ is from the observed time evolution.

\subsection{Summary for Evolving Wind}

To summarize the results of this section, the scalings we expect for the trapping radius and observed luminosity are
\be
    r_{\rm tr}/r_{\rm in} \propto 
    	\begin{cases}
    	1, & t\lesssim r_{\rm in}/v_w\\
		t, & r_{\rm in}/v_w\lesssim t
		\lesssim t'\\
		A(t), & t'\lesssim t \lesssim  t'A_{\rm max}^{1/\beta}\\
		1, & t\gtrsim t'A_{\rm max}^{1/\beta}.
  	\end{cases}
\ee
and
\be
    L_{\rm obs}/L_* \propto 
    	\begin{cases}
    	t^{-1/2}, & t\lesssim r_{\rm in}/v_w\\
		t^{-1/6}, & r_{\rm in}/v_w\lesssim t
		\lesssim t'\\
		A(t)^{-2/3}, & t'\lesssim t \lesssim  t'A_{\rm max}^{1/\beta}\\
		1, & t\gtrsim t'A_{\rm max}^{1/\beta},
  	\end{cases}
\ee
respectively. When $t'\lesssim r_{\rm in}/v_w$ (as in the blue and purple solutions in Figure \ref{fig:fallback_general2}) the second phase may be skipped entirely. Note that the timescales for each of these phases is approximate because of the time it takes to get to $r_{\rm tr}$ after the wind is launched, but this gives a sense of the scalings expected.

\section{Temperature Evolution}
\label{sec:temperature}

The above sections focus on the evolution of the luminosity of a wind-reprocessed transient, but another important observable is the temperature. At any depth the temperature is dominated by radiation so that
\be
    aT(r,t)^4 = \mathcal{E}(r,t),
\ee
where $a$ is the radiation constant. Below the trapping radius, this is set by the adiabatic cooling, so that
\be
    T(r,t)
        = \lp \frac{L_*[t_0(r,t)]}{4\pi r_{\rm in}v_wa}\rp^{1/4}
        \lp\frac{r}{r_{\rm in}} \rp^{-2/3}.
        \label{eq:trapped temperature}
\ee
Above the trapping radius, the luminosity is constant with depth. The energy density and temperature are then determined according to flux limited diffusion
\be
    L_{\rm obs}(t) = -\frac{4\pi r^2ac}{3\kappa_s\rho[t_0(r,t)]}
        \frac{\partial T(r,t)^4}{\partial r}.
    \label{eq:flux limited diffusion}
\ee
In practice, we simplify Equation (\ref{eq:flux limited diffusion}) when solving for the temperature distribution by dropping the $dr_{\rm tr}/dt$ in $L_{\rm obs}$. With the exception of the earliest rising phases, this introduces a less than $5\%$ error on $L_{\rm obs}$ and an even smaller error on the temperature estimate. Furthermore, we can solve Equation~(\ref{eq:flux limited diffusion}) analytically if we take $t_0\approx t$, resulting in
\be
    T(r,t) \approx
    \lb\frac{\kappa_sK(t)L_{\rm obs}(t)}{4\pi r^3ac} \rb^{1/4}.
\ee
Although this can provide a reasonable approximation for the temperature profile as long as $K$ is not changing too quickly with time, for the numerical examples considered later we solve Equation (\ref{eq:flux limited diffusion}) exactly.

If we set $\tau\approx 1$ using Equations (\ref{eq:tau1}) or (\ref{eq:tau2}), we can solve for the electron scattering photosphere. In general though, the effective temperature measured at this radius does not correspond to the observed color temperature. This is because the opacity for absorption is lower than that of electron scattering which dominates the wind. Thermalization requires that the wind is sufficiently optically thick that photon absorption can take place {\citep[also see the discussion in][]{Shen15}}. For an absorptive opacity $\kappa_a\ll \kappa_s$, we can define an effective opacity \citep{Rybicki86}
\be
    \kappa_{\rm eff} = (3\kappa_s\kappa_a)^{1/2},
\ee
and an associated effective optical depth
\be
    \tau_{\rm eff} = \int_r^{r_w}\kappa_{\rm eff} \rho dr,
\ee
where we note that $\kappa_a$ (and in turn $\kappa_{\rm eff}$) can be a function of the density and temperature in the wind. The condition $\tau_{\rm eff}\approx 1$ defines the color radius $r_c$. In general, $r_c$ can either be above or below $r_{\rm tr}$, so we next consider the resulting observed color temperature $T_{\rm obs}$ for each of these cases.

\subsection{Trapping-dominated Temperature}
\label{sec:trapping}

First, consider the case where $r_c<r_{\rm tr}$ (note this is different than shown in Figure \ref{fig:diagram}, where the trapping radius is interior to the color radius). In this instance, the photons are thermally coupled to the wind material out to the radius $r_c$, but then continue to be adiabatically cooled out to the radius $r_{\rm tr}$ due to advection. This means that the energy density of photons effectively evolves with depth as if they are coupled, until reaching the radius $r_{\rm tr}$ above which the photons diffuse out with few absorptions. Thus the observed temperature matches Equation (\ref{eq:trapped temperature}) evaluated at $r_{\rm tr}$,
\be
    T_{\rm obs}(t) =
    \lp \frac{L_*[t_0(r_{\rm tr},t)]}{4\pi r_{\rm in}^2v_w a} \rp^{1/4}
    \lp\frac{r_{\rm tr}(t)}{r_{\rm in}}\rp^{-2/3}.
\ee
Again, if we ignore the factor of $dr_{\rm tr}/dt$ for $L_{\rm obs}$, this can be rewritten as
\be
    T_{\rm obs} \approx
    \lp \frac{L_{\rm obs}}{4\pi r_{\rm tr}^2v_w a} \rp^{1/4}.
\ee
For different stages of the evolution, the temperature will roughly scale as
\be
    T_{\rm obs} \propto 
    	\begin{cases}
    	L_*(t)^{1/4}t^{-1/8}, & t\lesssim r_{\rm in}/v_w\\
		L_*(t)^{1/4}t^{-13/24}, & r_{\rm in}/v_w\lesssim t
		\lesssim t'\\
		L_*(t)^{1/4}A(t)^{-5/8}, & t'\lesssim t \lesssim  t'A_{\rm max}^{1/\beta}.
  	\end{cases}
  	\label{eq:tempobs_case1}
\ee
At sufficiently late times when $t\gtrsim t'A_{\rm max}^{1/\beta}$, the photons may not be able to thermalize at all and the observed temperature may better reflect the spectrum the photons were injected with.

\subsection{Thermalization-dominated Temperature}
\label{sec:thermalization}

The other case is when $r_c>r_{\rm tr}$ (as shown in Figure \ref{fig:diagram}). This is expected to occur for denser winds. Now even once the photons are no longer advected with the flow, they will continue to be thermalized with the wind material. The energy density of this material is set solving Equation (\ref{eq:flux limited diffusion}), so that
\be
    T_{\rm obs}(t)^4
    \approx -\int_{r_c(t)}^{r_w(t)} \frac{3\kappa_sK[t_0(r,t)]L_{\rm obs}(t)}{4\pi r^4 ac}dr.
\ee
For this case, it is more difficult to derive general scalings with time because $r_c$ will be evolving with time in a way that depends on the exact functional form of $\kappa_a$. This motivates us to consider some more specific examples in the next section.

\section{Specific Examples}
\label{sec:examples}

Since $r_c$ can evolve in more complicated ways than just simple scalings, here we consider some specific examples to better understand how the evolution proceeds. We still stick with power law scalings for both $L_*(t)$ and $K(t)$, namely
\be
    L_*(t) = L_{\rm max}(1+t/t')^{-\alpha},
\ee
and Equation (\ref{eq:power law k}) for $K(t)$. Different physical scenarios will result in different values of $L_{\rm max}$, $K_{\rm max}$, $t'$, $\alpha$, and $\beta$ as described next.

Furthermore, we need to consider a specific form for $\kappa_a$ for these calculations. The absorptive opacity can be a complicated function of density and temperature depending on the relative important of bound-bound, bound-free, and free-free interactions. For illustrative purposes, here we use a Kramer's opacity
\be
	\kappa_a = \kappa_0\rho T^{-3.5}\,{\rm cm^2\,g^{-1}}.
\ee
For the specific examples below, we use $\kappa_0=2\times10^{24}$ (assuming $\rho$ and $T$ are in cgs units), which is meant to mimic a bound-free opacity for roughly solar composition \citep{Hansen94}. In more detailed calculations, other opacity forms or tabulated opacities can be used. Although a helpful simplification is that as long as $r_c<r_{\rm tr}$, then the observed temperature no longer depends on the exact value of $\kappa_a$ as.

We note that in the following Sections \ref{sec:tde} and \ref{sec:bh birth}, we mostly summarize the main features of the solutions. The details of how we solve the system of equations for the radius, luminosity, and temperature evolution are presented in Appendix~\ref{sec:solving} and \ref{sec:specific examples}.

\subsection{Tidal Disruption Events}
\label{sec:tde}

For the first example, we consider the tidal disruption event (TDE) of a solar mass star by a supermassive black hole. There remains considerable uncertainty in where the powering originates from in such events, whether it be from a small amount of material fed into the BH \citep{Metzger16}, dissipation of stream self-interaction \citep{Piran15}, or secondary shocks \citep{2019arXiv190605865B}. In any scenario though it is generally thought that there is reprocessing of the emission from these sites because of the relatively low temperatures  ($\sim10^4\,{\rm K}$) measured from observations in comparison to what is expected from the emission regions ($\gtrsim10^5\,{\rm K}$).

For the disruption of a star with mass $M_*$  and radius $R_*$ by a black hole with mass $M_{\rm BH}$, the fallback of material falling a tidal disruption event scales roughly as
\be
    \dot{M}_{\rm fb} = \dot{M}_{\rm max}
    (1+t/t_{\rm fb})^{-5/3},
\ee
where
\be
    t_{\rm fb}
    \approx 41\lp \frac{M_{\rm BH}}{10^6\,M_\odot} \rp^{1/2}
    \lp \frac{M_*}{M_\odot} \rp^{-1} \lp\frac{R_*}{R_\odot} \rp^{3/2}
    {\rm days},
\ee
and
\be
    \dot{M}_{\rm max} &=& \frac{M_*}{3t_{\rm fb}}
    \nonumber
    \\
    &\approx& 2.9\lp \frac{M_{\rm BH}}{10^6\,M_\odot} \rp^{-1/2}
    \lp \frac{M_*}{M_\odot} \rp^{2}
    \lp\frac{R_*}{R_\odot} \rp^{-3/2}
    M_\odot\,{\rm yr^{-1}}.
    \nonumber
    \\
\ee
If this material is radiatively inefficient as it tries to accrete, only a fraction $\eta$ of this material can radiate its energy while the remaining energy goes into driving a wind. We follow \citet{Metzger16} and assume $\eta\ll 1$, which is also empirically supported by observed TDEs. We then take
\be
    L_* = \eta \dot{M}_{\rm fb}c^2,
\ee
and $K\approx \dot{M}_{\rm fb}/4\pi v_w$ with $v_w\approx10^9\,{\rm cm\,s^{-1}}$ motivated by both observations of TDEs and theoretical expectations for the escape velocity. For the inner radius, we use $r_{\rm in}\approx 100r_g$, where $r_g=GM_{\rm BH}/c^2=1.5\times10^{11}(M_{\rm BH}/10^6\,M_\odot)\,\rm cm$.

\begin{figure}
\includegraphics[width=0.48\textwidth,trim=0.0cm 0.0cm 0.0cm 0.5cm]{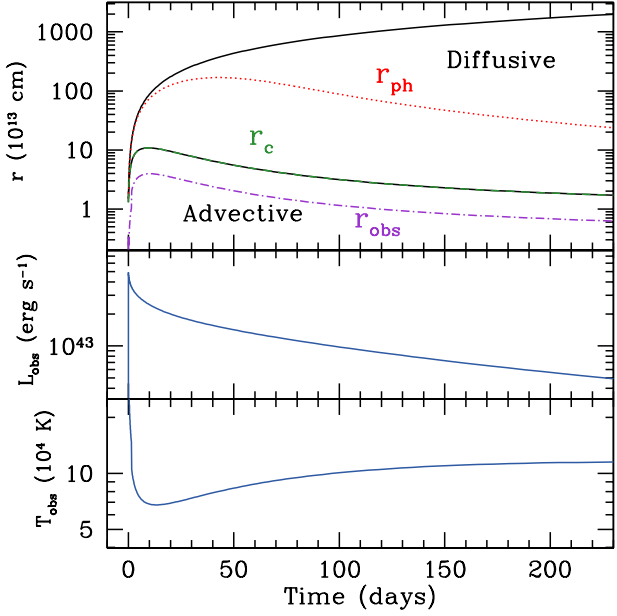}
\caption{Evolution of the TDE of an $M_*=M_\odot$ star by an $M_{\rm BH}=10^6\,M_\odot$ using the framework developed here. For the injected luminosity we set $\eta=10^{-3}$. In the top panel, the upper black line is $r_w$ and the lower black line is $r_{\rm tr}$. These divide the diffusive and advective regions of the flow as labeled. The other key lines are the scattering photosphere $r_{\rm ph}$ (dotted red line), the radius where the temperature is determined $r_c$ (dashed green line), and inferred observed radius (dot-dashed purple line). The color depth is always below the trapping radius in this case and thus the temperature is determined at the trapping radius. The middle and bottom panels show the observed bolometric luminosity and temperature, respectively.}
\label{fig:tde}
\end{figure}

An example solution is presented in Figure \ref{fig:tde}. In addition to the other key radii that are described above, we also plot the scattering photosphere $r_{\rm ph}(t)$, defined as
\be
    \tau[r_{\rm ph}(t)]
    = \int_{r_{\rm ph}(t)}^{r_w}
     \frac{\kappa_sK[t_0(r,t)]}{r^2} dr = 1,
\ee
and the observationally inferred radius, defined as
\be
    r_{\rm obs}
    \equiv \lp \frac{L_{\rm obs}}{4\pi \sigma_{\rm SB}T_{\rm obs}^4}\rp^{1/2}.
\ee
This latter radius is what an observer would infer from assuming that the TDE emission is simply black body. We can see that this is actually much smaller than any of the other key radii associated with this event, because the thermalization is so weak in the outer layers of the wind (due to $\kappa_a\ll\kappa_s$). The rise-segment of the lightcurve (for $t\lesssim t_{\rm fb}$) should be smoother if a more realistic fallback rate from simulations is adopted \citep[e.g.,][]{2009MNRAS.392..332L, 2013ApJ...767...25G}.

This general framework is able to replicate the main features of TDEs, namely, a photosphere that recedes with time, a falling luminosity in the range of $\sim10^{43}-10^{44}\,{\rm erg\,s^{-1}}$, and a roughly constant temperature in the range of $\sim10^4-10^5\,{\rm K}$. Our model mostly follows the results of \citet{Metzger16}, but with a detailed consideration of what sets the color temperature. \citet{Roth16} nicely present through analytic and numerical arguments where the color temperature is set (referred to as the ``optical continuum photosphere'' in this work), but our study differs in that we follow the trapping radius. Since we find that the color depth is below the trapping depth for this toy model, this conveniently means that the details of $\kappa_a$ do not matter as much for setting the observed temperature. However, if one includes a detailed treatment of bound-free and bound-bound absorption (with appropriate line broadening), the color depth may be above the trapping depth, and in that case, the observed color temperature is lower than obtained here \citep{2020MNRAS.492..686L}.

This model demonstrates that the photosphere inferred via observations \citep[e.g.,][]{Holoien16} and fitting techniques \citep[e.g.,][]{Mockler19} that assume black body emission are not fitting for the true emission radius because the wind is highly scattering dominated. Making the approximation that $r_{\rm tr}\ll r_w$ and that $K(t)$ is not changing too quickly with time, we can estimate
\be
    r_{\rm obs}\approx \lp \frac{4v_w}{c}\rp^{1/2} r_{\rm tr}
    \approx \lp \frac{4v_w}{c}\rp^{1/2}
    \frac{\kappa_s\dot{M}}{4\pi c}.
    \label{eq:r_inf}
\ee
Thus one may be able to use this inferred radius to learn more about the wind surrounding the event.

\subsection{Stellar Mass Black Hole Formation}
\label{sec:bh birth}

The second example we consider the fallback of material onto a newly born black hole following unsuccessful core collapse, similar to the scenario envisioned by \citet{Kashiyama15}. The basic picture is of a massive star that collapses in a failed supernova to become a black hole. With sufficient angular momentum, the fallback material produces an accretion disk around the newly born black hole. Such disks generally produce super-Eddington accretion rates which drive strong disk winds. Accretion onto the black hole illuminates these winds, leading to a fast blue transient.

\begin{figure}
\includegraphics[width=0.48\textwidth,trim=0.0cm 0.0cm 0.0cm 0.5cm]{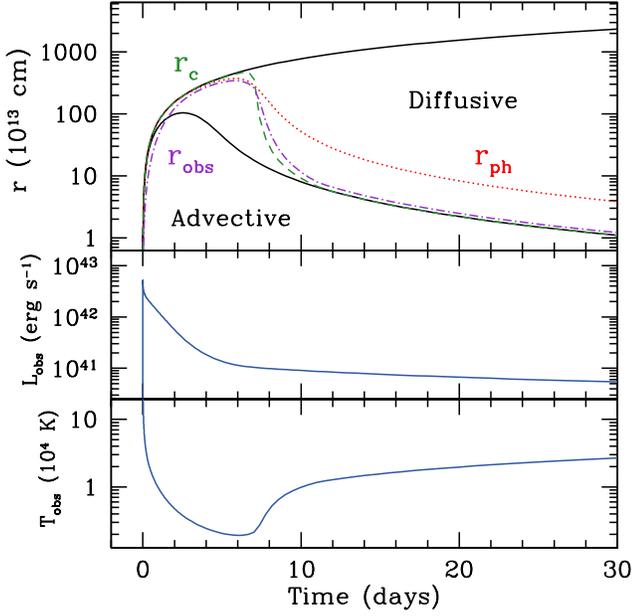}
\caption{The same as Figure \ref{fig:tde}, but for a stellar mass BH forming event. Here, initially $r_c$ is beyond the trapping radius, but then becomes equal to $r_{\rm tr}$ at $\sim10\,{\rm days}$.}
\label{fig:bh_formation}
\end{figure}

In this case, we again expect that the fallback rate scales as $t^{-5/3}$ as shown in the work of \citet{Dexter13}. The relevant timescale is now the fallback time or
\be
    t_{\rm fb} &\approx& \pi (R_*^3/8GM_{\rm BH})^{1/2}
    \nonumber
    \\
    &\approx& 7 \lp\frac{M_{\rm BH}}{7\,M_\odot} \rp^{-1/2}
    \lp \frac{R_*}{10\,R_\odot}\rp^{3/2}{\rm hr},
\ee
where we use a BH mass similar to the peak of the galactic black hole mass distribution \citep{ozel10} and a radius that would be appropriate for a Wolf-Rayet or blue supergiant star. The corresponding accretion rate for a disk with mass $M_d$ can be estimated as
\be
    \dot{M}_d &\approx& M_d/t_{\rm fb}
    \nonumber
    \\
    &\approx& 5\times10^{-5}
    \lp\frac{M_d}{M_\odot} \rp
    \lp\frac{M_{\rm BH}}{7\,M_\odot} \rp^{1/2}
    \lp \frac{R_*}{10\,R_\odot}\rp^{-3/2}M_\odot\,{\rm s^{-1}}.
    \nonumber
    \\
\ee
For simplicity, we assume that the fallback is the rate limiting step for feeding the BH, rather than the viscous time of the disk. This is supported by simulations of accretion of low angular gas \citep[e.g.,][]{Proga03a,Proga03b}, but in future work one could also track the viscous evolution of the disk including a fallback term using a simple disk model to better track the time dependent accretion rate \citep[e.g.,][]{Metzger08}.

Since this accretion rate is orders of magnitude greater than the Eddington accretion rate for this black hole, we again assume that the majority of this mass is blown in a wind while merely a fraction $\eta\ll1$ is accreted by the BH just as in our consideration of TDEs. The typical launching radius of this wind is the characteristic radius of the disk, which depends on the angular momentum of the star. Since this can vary depending on the mass loss history of a given star, we parameterize this radius with the factor $f_d$, so that
\be
    r_d \approx \frac{2GM_{\rm BH}}{c^2}f_d
    \approx 2\times10^7 \lp\frac{f_d}{10} \rp
    \lp\frac{M_{\rm BH}}{7\,M_\odot}  \rp{\rm cm}.
\ee
This implies a relatively high launching velocity,
\be
    v_w \approx \lp \frac{2GM_{\rm BH}}{r_d}\rp^{1/2}
    \approx 0.3 \lp\frac{f_d}{10}  \rp^{-1/2}c.
\ee
As in the TDE case, we assume that a fraction of the material $\eta$ is able to accrete onto the newly born BH, which produces a luminosity of $L_d\approx \eta \dot{M}_dc^2$ that illuminates these winds.

The resulting luminosity evolution is shown in Figure \ref{fig:bh_formation}, where we have used the above parameters along with $\eta=10^{-2}$ (see Appendix \ref{sec:solving} for more details about these solutions). The main difference in comparison to the TDE case is that the winds are now much denser due to the high $\dot{M}_d$. This causes the color radius to be above the trapping radius at early times (the case described in Section \ref{sec:thermalization}). Nevertheless, the general observed features are largely similar, with an inferred emission radius that moves to smaller radii and a rather constant or slightly increasing observed temperature. {The photospheric radius evolution we find is qualitatively different from the models studied by \citet{Kashiyama15}. In their case, the accretion onto the BH dramatically drops once the stellar surface falls in. This results in ejecta concentrated in a narrow radius and a photosphere that moves outward. Here we use a fallback rate that  falls as $t^{-5/3}$ at late times. Such a scaling is applicable to a scenario where a low energy explosion expands the star but is ultimately unsuccessful in unbinding it \citep[e.g.,][]{Dexter13}. }

\subsection{Other Scenarios}

The above examples are meant to provide some sense of the range of systems that can be addressed with the framework presented here. There are many other scenarios where this work could be applied in future investigations.

{\it Wind Collision.} One example would be the collision of a wind with circumstellar material (CSM). As the wind with mass loss rate $\dot{M}$ collides with the CSM, some fraction $\eta$ of its kinetic energy would be converted to a luminosity $L_*=\eta \dot{M}v_w^2/2$. {Here $\eta$ roughly corresponds to the fraction of solid angle subtended by the CSM. As the collision occurs, the shocked wind moves more slowly than that unimpeded wind. The radiation produced in the shocked regions must then diffuse through the unimpeded wind to get to the observer.} Although our framework is one-dimensional, it would still be fairly accurate for this case as long as the majority of the wind gets past the CSM and the emission is not too viewing angle dependent.

{\it Magnetar Formation.} Another possible scenario is the formation of a highly magnetized neutron star (magnetar). The basic picture is that following the merger of two neutron stars, it is likely in many cases that the result will either be a neutron star or at least a remnant that can hold off collapsing to a BH for a short while due to its high spin. The strong differential rotation of this process can generate large magnetic field ($B\gtrsim10^{15}\,{\rm G}$; \citealp{Price06,Zrake13,Duncan92}), which produces a high luminosity of $\sim10^{48}\,{\rm erg\,s^{-1}}$ from the spin down. At the same time, this remnant would be surrounded by a $\sim0.1\,M_\odot$ disk of material that produces winds through its low radiative efficiency and heating via neutrinos. Such a scenario has been considered for a magnatar with supernova-like ejecta \citep{Metzger14}, but the emission may be qualitatively different if the environment was dominated by winds \citep[e.g.,][]{Dessart09,Fernandez13}.

{\it Nuclear Heating.} Finally, another energy source we have not considered in the above examples is nuclear heating. In cases were a white dwarf explosion does not successfully unbind the star, the remnant may produce winds driven by leftover radioactive material \citep{Shen17}. Applying this framework to such a scenario may help to better understand the Type Iax supernovae \citep{Foley13} that have been hypothesized to be these failed explosions \citep{Foley14}. Although at sufficiently late times, these winds may be cool enough to produce dust \citep{Fox16,Foley16}, which is not accounted for in this work.

\section{Interpreting AT2018cow}
\label{sec:18cow}

The fast, blue transient AT2018cow \citep{Prentice18, 2019ApJ...871...73H} showed a power-law declining luminosity, a receding inferred photosphere, roughly thermal spectra, and radio/X-ray emission indicative of some sort of underlying power source. Despite extensive multiband observations of this event \citep{Kuin19,Margutti19,Perley19,Prentice18,2019ApJ...871...73H, RiveraSandoval18}, there is no agreed upon explanation for its origin, with ideas including a TDE by an intermediate mass BH \citep{Perley19}, collapse of a massive star to produce a BH \citep{Quataert19}, magnetar creation \citep{Margutti19}, electron capture of a merged white dwarf \citep{Lyutikov19}, shocked disk interaction buried within a supernova \citep{Margutti19}, and a common envelope with jets \citep{Soker19}. Nevertheless, many of these features show similarities to what we would expect for a wind-reprocessed transient, in particular the declining radius. Whether or not this is the ultimate explanation for AT2018cow, we can at least investigate what our model would imply about the physical parameters associated with this transient.

If we assume that AT2018cow follows the case where the temperature is determined at the trapping radius, we use Equation (\ref{eq:r_inf}) to estimate the wind mass loss rate at any give time
\be
    \dot{M} \approx
    \frac{4\pi cr_{\rm obs}}{\kappa_s}\lp\frac{4v_w}{c}\rp^{1/2}.
    \label{eq:mdot obs}
\ee
This ignores the time it takes to travel from the inner to the trapping radius, but this is at least good enough to get a rough idea of what $\dot{M}$ should be. Next, the observed optical luminosity and radius can be used to estimate what underlying luminosity was injected into the wind
\be
    L_* \approx L_{\rm obs}
        \lp\frac{r_{\rm tr}}{r_{\rm in}}\rp^{2/3}
        \approx  L_{\rm obs}
        \lp\frac{r_{\rm obs}}{r_{\rm in}}\rp^{2/3}
        \lp \frac{c}{4v_w}\rp^{1/3}.
        \label{eq:lstar obs}
\ee
Given $L_{\rm obs}$ and $r_{\rm obs}$ from the observations, we can derive what $L_*$ and $\dot{M}$ should be. The only unknowns are $v_w$ and $r_{\rm in}$.

\begin{figure}
\includegraphics[width=0.48\textwidth,trim=0.0cm 0.0cm 0.0cm 0.5cm]{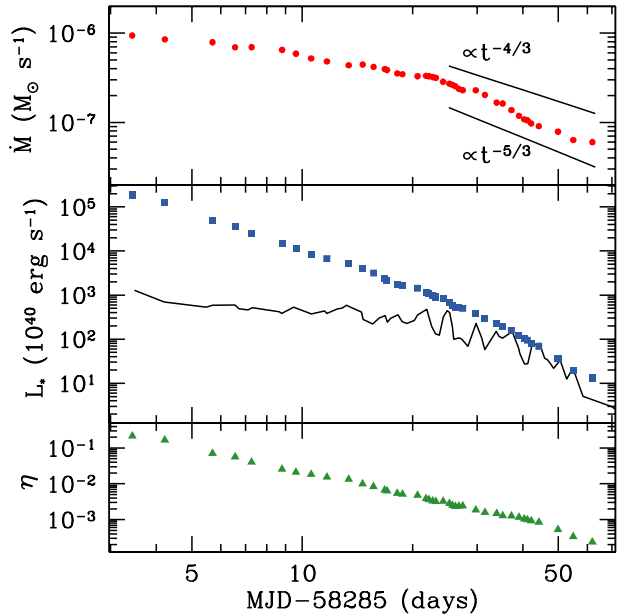}
\caption{Inferred $\dot{M}$ (red circles) and $L_*$ (blue squares) for AT2018cow, assuming that it is wind-reprocessed with $v_w=0.1c$ and $r_{\rm in}=10^{14}\,{\rm cm}$. {The bottom panel plots $\eta=L_*/(\dot{M}v_w^2/2)$ (green triangles), the efficiency inferred assuming that the powering luminosity is related to the wind mass loss rate.} The late time mass rate asymptotes to a power law. An index in the range of $\approx -4/3$ to $-5/3$ suggests that a disk wind or fallback scenario could explain this reprocessing material. At late times, $L_*$ is similar to the observed X-ray luminosity $L_X$ \citep[black line;][]{Margutti19}, suggesting that the optical emission results from these reprocessed X-rays. At early times $L_*>L_X$, which is expected if the X-rays are highly obscured. {The efficiency $\eta$ is required to be $\approx 0.1-0.2$ early on and falls to $\lesssim10^{-3}$ after about 50 days. We discuss potential reasons for this behavior in the text.}}
\label{fig:at2018cow}
\end{figure}

Figure \ref{fig:at2018cow} shows the results of this analysis for $v_w=0.1c$ and $r_{\rm in}=10^{14}\,{\rm cm}$. The wind velocity is motivated by the rapid rise of the optically inferred photospheric radius in the first few days \citep{Perley19}, as well as the shock speed inferred from the evolution of the flux at the synchrotron self-absorption frequency \citep{2019ApJ...871...73H}. The upper panel shows the inferred $\dot{M}$ (red circles), which shows a wind loss rate in the range of $\dot{M}\approx 10^{-8}-10^{-6}\,M_\odot\,{\rm s^{-1}}$ with a broken power-law evolution. The range of power laws that fit the late time $\dot{M}$ evolution are suggestive of a wind, either in the case of radiatively inefficient disk wind ($\dot{M}\sim t^{-4/3}$) or fallback accretion that is blown into a wind ($\dot{M}\sim t^{-5/3}$). 

The middle panel shows the inferred $L_*$ (blue squares). In comparison, we include the soft ($0.3-10\,{\rm keV}$) X-ray luminosity observed \citep[black line;][]{Margutti19}. {The similarity of the inferred injection luminosity $L_*$ and the rough time evolution of the X-rays (if the X-rays are considered in a time-averaged sense) supports the suggestion of \citet{Margutti19} that the soft X-rays are reprocessed to produce the optical emission. The normalization of $L_*$ does depend on the chosen value of $r_{\rm in}$, but this comparison shows that $r_{\rm in}\approx10^{14}\,{\rm cm}$ gives a reasonable explanation for this event. At early times, $L_*$ is much greater than the observed X-ray luminosity. This is consistent with the idea of \citet{Margutti19} that the X-rays are initially partially obscured but then mostly observed at later times.}

{In the lower panel of Figure \ref{fig:at2018cow}, we plot the inferred efficiency $\eta=L_*/(\dot{M}v_w^2/2)$  (green triangles) under the assumption that the wind mass loss rate is connected to the powering luminosity. This connection is not in general required because $L_*$ could be from some underlying engine, but it is interesting to consider $\eta$ and what it might imply about AT2018cow. We see that $\eta\approx0.1$ at early times and drops to below $\approx10^{-3}$ at late times. The decrease of $\eta$ is intriguing and we offer two speculations for this behavior. One possibility is that the X-rays are produced by a wind interacting with an equatorial torus of material. As the torus is eroded (only the densest clumps survives the shocks at late time), the efficiency of converting the kinetic energy of the wind into X-rays decreases. A second possibility is that $v_w$ is not constant as assumed and is instead dropping with time. In the case of a viscously spreading disk (a compact inner disk), $v_w$ will decrease as the disk radius increases according to the local escape speed. Future more detailed modeling is needed to determine if either of these scenarios are possible or if another explanation could explain for these trends we infer.}

\section{Conclusions and Discussion}
\label{sec:conclusions}

We considered the general properties of a transient that is being reprocessed by an optically thick wind. We separately studied the cases of a steady wind and a wind that is changing with time. We discussed how these winds will likely be scattering dominated and explore how this impacts the observed temperature evolution. {This framework is applied to two specific cases, TDEs (where $A_{\rm max}$ is small and $t'$ is large) and stellar mass BH formation (where $A_{\rm max}$ is large and $t'$ is small) to provide more concrete examples of how reprocessing can work.}

Given a $K(t)$ and $L_*(t)$ for some specific scenario, the methods presented in Section \ref{sec:evolving} can be used to solve for observed luminosity. A wide range of different potential applications are described in Section~\ref{sec:examples}, and even among these there are different variations that would be interesting to explore. This framework may be especially helpful when evaluating numerical models where the hydrodynamics are solved for but the radiative transfer is not included because it is deemed too expensive. This can be done by using the numerical output to find $K(t)$ and $L_*(t)$ and then solving the equations presented in Section~\ref{sec:basic} numerically (including tabulated $\kappa_a$ and $\kappa_s$).

Finally, we discussed the transient AT2018cow in the context of a wind-reprocessed framework. At a basic level, this example provides a template for how to approach other transients in the future and assess whether a wind-reprocessed model is applicable. One can use Equations~(\ref{eq:mdot obs}) and (\ref{eq:lstar obs}) to estimate the evolution of $\dot{M}$ and $L_*$ needed to make a wind-reprocessed model work. These in turn can be used to judge whether such a model is physically reasonable.

An important diagnostic which indicates that such a model should be considered is {\it an inward propagating radius.} This is a classic property of TDEs and also something that made AT2018cow stand out in comparison to many other transients. Unfortunately, for many of the interesting fast transients that have been discovered the radius evolution has not been followed \citep[e.g.,][]{Ho2020}, but it should be a priority to measure this property in the future. A decreasing radius has even been inferred for some seemingly normal core-collapse SNe \citep[e.g., SN 2018bbc,][]{Karamehmetoglu19}, suggesting broader application of this model.

Applying this model to AT2018cow, the two main conclusions were that (1) the $\dot{M}$ evolves as a power law, which is suggestive of a disk-wind or fallback scenario, and (2)~that $L_*$ is similar to the X-rays observed from AT2018cow, which fits with a picture where these X-rays are reprocessed to produce the optical emission. Beyond these basic properties, the wind-reprocessed model for AT2018cow is fairly agnostic to the details of the source of the X-rays, but it is interesting to speculate. The time dependence of $L_*$ is much steeper than would be expected for millisecond magnetar spin down (which would give $L_*\sim t^{-2}$) or fallback from a TDE ($L_*\sim t^{-5/3}$). This makes it difficult to explain with such a picture unless the efficiency for producing the X-ray emission is changing dramatically with time.

{Perhaps more attractive is X-rays produced from shock interaction as described by \citet{Margutti19} or even by \citet{Andrews18} in the context of iPFT14hls. In this picture, a centrally launched wind interacts with pre-explosion equatorial material (similar to the ejecta-equatorial ring interaction in SN 1987A), producing the observed X-rays. The rough scale of $r_{\rm in}\sim10^{14}\,{\rm cm}$ needed for $L_*$ to be comparable to the observed X-ray luminosity should depend on the conditions of the pre-existing equatorial material and should be checked against expectations for equatorial outflows before explosion. The variability of the X-rays would naturally be explained by interaction with clumps of matter, which would be more difficult to understand in models where the powering is more like an engine. The winds would originate from a long-lived disk accreting onto the central compact object and have a time dependence dictated by either fallback material or radiatively inefficient viscous evolution.}

More work should be done to understand the three dimensional structure of wind material in AT2018cow. This is outside the scope of this current work since our main goal is to present the basic framework of wind-reprocessed transients. Nevertheless, given the mass loss rates estimated here, one should check whether the X-ray evolution (both in soft and hard bands) makes sense. Such an investigation would help constrain the covering fraction of the equatorial material and the viewing angle of the observer. It would also hopefully provide a better understanding of the stellar progenitor required to make AT2018cow.

\acknowledgments
{We thank the anonymous referee for a careful reading of our work and helpful feedback. We thank Sterl Phinney and Anna Ho for useful discussions on AT2018cow.} W.L. is supported by the David and Ellen Lee Fellowship at Caltech.

\begin{appendix}
\counterwithin{figure}{section}

\section{Numerically Solving for an Evolving Wind}
\label{sec:solving}

Here we derive the equations for an evolving wind using dimensionless variables to help with finding numerical solutions. First, we define dimensionless radial and time variables
\be
    \chi \equiv r/r_{\rm in}, \hspace{0.25cm} \xi \equiv v_wt/r_{\rm in}.
\ee
Then the relation for the launching time from Equation (\ref{eq:launchingtime}) becomes
\be
    \xi = \xi_0 + \chi -1,
\ee
where $\xi_0=v_wt_0/r_{\rm in}$. We define a new optical depth
\be
    \tilde{\tau}(\chi,\xi) = \tau(r,t)v_w/c
        = \int_\chi^{\chi_w} \frac{A[\xi_0(\chi,\xi)]}{\chi^2}d\chi,
    \label{eq:dimensionlesstau}
\ee
where
\be
    \chi_w = 1+\xi,
\ee
and the dimensionless wind parameter is
\be
    A(\xi) \equiv \kappa_s K(\xi) v_w/r_{\rm in}c.
\ee
The condition for finding the trapping radius given by Equation (\ref{eq:rtr_condition}) is simplified to be
\be
    \tilde{\tau}(\chi_{\rm tr},\xi) \chi_{\rm tr} (1+\xi-\chi_{\rm tr})
        = \chi_w(\xi-\xi_0).
    \label{eq:dimensionlesstrapping}
\ee
From here we find $\chi_{\rm tr}$ for a given $\xi$ via iteration. The steps are as follows. (1) A trial $\chi_{\rm tr}$ is chosen between $1$ and $\chi_w$. (2) We integrate Equation (\ref{eq:dimensionlesstau}) to find $\tilde{\tau}(\chi_{\rm tr},\xi)$. (3) This is substituted into Equation (\ref{eq:dimensionlesstrapping}). (4) If the left-hand side is too big, then we need to choose a smaller $\chi_{\rm tr}$, and the converse if the left-hand side is too small. With a new trial $\chi_{\rm tr}$, we go back to step (2) until we have converged on the correct value of $\chi_{\rm tr}$. Once the trapping radius is found, we use
\be
    L_{\rm obs}(\xi) = L_*(\xi_0)\chi_{\rm tr}^{-2/3}
    \lp 1- \frac{d\chi_{\rm tr}}{d\xi} \rp,
\ee
to find the observed luminosity as a function of time.

\section{Dimensionless Variables for Specific Examples}
\label{sec:specific examples}

For the cases where the wind loss rate is evolving with time as a power law as considered in Sections~\ref{sec:power law} and \ref{sec:examples}, we use a dimensionless wind parameter
\be
    A(\xi) = A_{\rm max}(1+\xi/\xi')^{-\beta},
\ee
where $\xi'\equiv v_wt'/r_{\rm in}$. If we set $L_*=\eta \dot{M}c^2$, $L_*$ in terms of $A$ is
\be
    L_* =  4\pi r_{\rm in} \eta A c^3/\kappa_s.
\ee
For a TDE as discussed in Section \ref{sec:tde},
\be
    A_{\rm max} = \frac{\kappa_s\dot{M}_{\rm max}}{4\pi r_{\rm in}c}
    = 16.7 \lp \frac{M_{\rm BH}}{10^6\,M_\odot} \rp^{-1/2}
    \lp \frac{M_*}{M_\odot} \rp^{2}
    \lp\frac{R_*}{R_\odot} \rp^{-3/2}
    \lp\frac{r_{\rm in}}{10^{13}\,{\rm cm}} \rp^{-1},
\ee
and
\be
    \xi' = \frac{v_wt_{\rm fb}}{r_{\rm in}}
    = 3.5\times10^2 
    \lp \frac{M_{\rm BH}}{10^6\,M_\odot} \rp^{1/2}
    \lp \frac{M_*}{M_\odot} \rp^{-1} \lp\frac{R_*}{R_\odot} \rp^{3/2}
     \lp\frac{v_w}{10^{9}\,{\rm cm\,s^{-1}}} \rp
     \lp\frac{r_{\rm in}}{10^{13}\,{\rm cm}} \rp^{-1}.
\ee
For fallback stellar mass BH formation event as discussed in Section \ref{sec:bh birth}, we take $r_{\rm in}\approx r_d$, so that 
\be
    A_{\rm max} = \frac{\kappa_s\dot{M}_d}{4\pi r_dc}
    = 1.3\times10^9
    \lp\frac{f_d}{10} \rp^{-1}
    \lp\frac{M_d}{M_\odot} \rp
    \lp\frac{M_{\rm BH}}{7\,M_\odot} \rp^{-1/2}
    \lp \frac{R_*}{10\,R_\odot}\rp^{-3/2},
\ee
where we use $\kappa_s\approx0.1\,{\rm cm^2\,g^{-1}}$, as would be appropriate for partially ionized hydrogen-deficient material, and
\be
    \xi' = \frac{v_wt_{\rm fb}}{r_d}
    = 1.1\times10^7 
    \lp\frac{f_d}{10} \rp^{-3/2}
    \lp\frac{M_{\rm BH}}{7\,M_\odot} \rp^{-3/2}
    \lp \frac{R_*}{10\,R_\odot}\rp^{3/2}.
\ee
Thus the TDE and BH formation cases span the parameter range from low to high values of $A_{\rm max}$ and $\xi'$.

\end{appendix}

\bibliographystyle{yahapj}

\end{document}